# MULTIPLE SURFACE-PLASMON-POLARITON WAVES GUIDED JOINTLY BY A METAL AND A HYPERBOLIC, DIELECTRIC, STRUCTURALLY CHIRAL MATERIAL

Akhlesh LAKHTAKIA* and Muhammad FARYAD

Department of Engineering Science and Mechanics,
Pennsylvania State University, University Park, PA 16802, USA

*E-mail: akhlesh@psu.edu

The planar interface of a metal and a hyperbolic, dielectric, structurally chiral material can support the propagation of none, one, or multiple surface-plasmon-polariton (SPP) waves, at a specified frequency and along a specified direction in the interface plane. When multiple SPP waves can exist, they differ in phase speed, propagation length, degree of localization to the interface, and spatial profiles of the associated electromagnetic fields.

*Key words*: chiral sculptured thin film, surface multiplasmonics, hyperbolic materials

## 1. INTRODUCTION

The emerging area of surface multiplasmonics encompasses the excitation and the propagation of multiple surface-plasmon-polariton (SPP) waves—all of the same frequency, but different phase speed, propagation length, degree of localization to the interface, and spatial profiles of fields—guided by the planar interface of two partnering materials, one of which allows the existence of quasi-particles called plasmons in it whereas the other is polarizable at the frequency of interest [1]. The plasmonic partner is usually a homogeneous metal. The polarizable partner is a dielectric material that can be either isotropic [2] or anisotropic [3], but it must be periodically nonhomogeneous in the direction normal to the interface [4]. The multiplicity has been experimentally confirmed and even put to use in an optical sensor [5]. If the dielectric partner is homogeneous, then only one SPP wave is possible at a specific frequency [6, 7, 8, 9].

Virtually all of the literature on SPP waves is based on the assumption that the real part of the relative permittivity dyadic $\underline{\underline{\varepsilon}}_r$ of the dielectric partner is positive definite, i.e., all three eigenvalues of $\underline{\underline{\varepsilon}}_r$ have positive real parts [10]. The real part of the relative permittivity $\varepsilon_m$ of the plasmonic partner is negative. What would happen if $Re(\underline{\underline{\varepsilon}}_r)$ were indefinite, i.e., if either one or two of its eigenvalues had negative real parts, the remaining having positive real parts? Such materials are nowadays called hyperbolic materials [11, 12]. Examples exist in nature [13, 14, 15] and have also been manufactured [16, 17].

In the context of surface multiplasmonics, a hyperbolic material partnering a metal [$Re(\varepsilon_m) < 0$] must be periodically nonhomogeneous normal to the interface. A candidate is the hyperbolic, dielectric, structurally chiral material [18] described by the frequency-domain constitutive relations

$$\left.\begin{aligned}
\boldsymbol{D}(\boldsymbol{r}) &= \varepsilon_0 \underline{\underline{\varepsilon}}_r(z) \cdot \boldsymbol{E}(\boldsymbol{r}) \\
&= \varepsilon_0 \underline{\underline{S}}_z(z) \cdot \underline{\underline{S}}_y(\chi) \cdot \left(\varepsilon_a \hat{u}_z \hat{u}_z + \varepsilon_b \hat{u}_x \hat{u}_x + \varepsilon_c \hat{u}_y \hat{u}_y\right) \cdot \underline{\underline{S}}_y^{-1}(\chi) \cdot \underline{\underline{S}}_z^{-1}(z) \cdot \boldsymbol{E}(\boldsymbol{r}) \\
\boldsymbol{B}(\boldsymbol{r}) &= \mu_0 \boldsymbol{H}(\boldsymbol{r})
\end{aligned}\right\} \tag{1}$$

where the direction of nonhomogeneity is parallel to the $z$ axis; $\mu_0$ and $\varepsilon_0$ are the permeability and permittivity of free space; the periodic nonhomogeneity is expressed through the rotation dyadic

$$\underline{\underline{S}}_z(z) = \hat{u}_z \hat{u}_z + \left(\hat{u}_x \hat{u}_x + \hat{u}_y \hat{u}_y\right)\cos\left(\frac{h\pi z}{\Omega}\right) + \left(\hat{u}_y \hat{u}_x - \hat{u}_x \hat{u}_y\right)\sin\left(\frac{h\pi z}{\Omega}\right) \tag{2}$$

with $2\Omega$ as the period and either $h = +1$ for structural right-handedness or $h = -1$ for structural left-handedness; the dyadic

$$\underline{S}_y(\chi) = \hat{u}_y\hat{u}_y + (\hat{u}_x\hat{u}_x + \hat{u}_z\hat{u}_z)\cos\chi + (\hat{u}_z\hat{u}_x - \hat{u}_x\hat{u}_z)\sin\chi \qquad (3)$$

contains the tilt angle $\chi \in [0, \pi/2]$ with respect to the $xy$ plane; $\varepsilon_a$, $\varepsilon_b$, and $\varepsilon_c$ are the three $z$-independent eigenvalues of $\underline{\varepsilon}_r(z)$; and either one or two of these three eigenvalues have negative real parts but the remainder do not.

With an $exp(-i\omega t)$ time dependence, the canonical problem of SPP-wave propagation is as follows [19]: The half space $z < 0$ is occupied by a metal with relative permittivity scalar $\varepsilon_m$, the half space $z > 0$ by the hyperbolic, dielectric, structurally chiral material described by Eqs. (1)-(3), and the electromagnetic field phasors everywhere can be written as

$$\left.\begin{array}{l} \boldsymbol{E}(r) = \boldsymbol{e}(z)\exp[iq(x\cos\psi + y\sin\psi)] \\ \boldsymbol{H}(r) = \boldsymbol{h}(z)\exp[iq(x\cos\psi + y\sin\psi)] \end{array}\right\} \qquad (4)$$

with the unknown SPP wavenumber q and unknown functions $\boldsymbol{e}(z)$ and $\boldsymbol{h}(z)$. The angle $\psi \in [0, 2\pi)$ denotes the direction of propagation in the $xy$ plane. As the procedure to solve a dispersion equation for q, and then determine $\boldsymbol{e}(z)$ and $\boldsymbol{h}(z)$, for a specific $\psi$ has been described elsewhere in detail [1, 19], it is not repeated here.

## 2. NUMERICAL RESULTS AND DISCUSSION

For illustrative results, we set $\varepsilon_m = -56 + i21$ (aluminum at free-space wavelength $\lambda_0 = 633$ nm), $\varepsilon_a = 2.26(1 + i\delta)$, $\varepsilon_b = 3.46(-1 + i\delta)$, $\varepsilon_c = 2.78(1 + i\delta)$, $\delta = 0.001$, $h = +1$, $\Omega = 135$ nm, $\chi = \pi/6$, and $\psi = 0$. Over the spectral regime $\lambda_0 \in [600, 700]$ nm, we found either one or two solutions $q$ of the SPP-wave dispersion equation. Our search was confined to the regime $\mathrm{Re}(q)/k_0 \in (1, 4]$, where $k_0 = 2\pi/\lambda_0$ is the free-space wave number, and we incremented $\lambda_0$ in units of 5 nm. As shown in Fig. 1, the solutions can be organized on two branches. Two SPP waves that differ in phase speed $\omega/\mathrm{Re}(q)$ and propagation length $1/\mathrm{Im}(q)$ can propagate along the $x$ axis in the interface plane for $\lambda_0 \in [600, 661]$ nm, but only one such SPP wave can exist for $\lambda_0 \in [662, 700]$ nm. SPP waves on the longer branch have a phase speed $\sim 0.38c_0$, where $c_0$ is the speed of light in free space, while SPP waves on the shorter branch have a phase speed $\sim 0.82c_0$. The propagation lengths of SPP waves on the longer branch are $\sim 4.2$ μm, while those on the shorter branch have propagation lengths ranging from $\sim 47$ μm to $\sim 94$ μm.

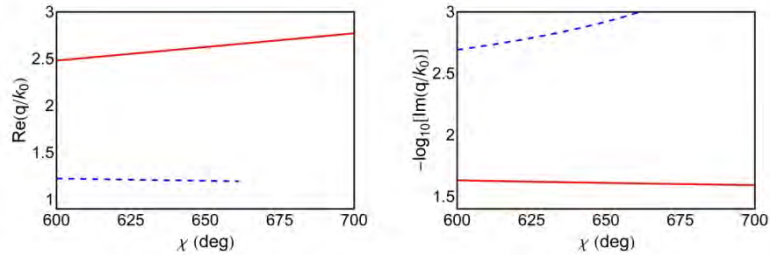

Fig. 1 Real and imaginary parts of $q$ as functions of $\lambda_0$ when $\psi = 0$, $\varepsilon_m = -56 + i21$, $\varepsilon_a = 2.26(1 + i\delta)$, $\varepsilon_b = 3.46(-1 + i\delta)$, $\varepsilon_c = 2.78(1 + i\delta)$, $\delta = 0.001$, $h = +1$, $\Omega = 135$ nm, and $\chi = 30°$.

Not only are the SPP waves on the two branches in Fig. 1 dissimilar in phase speed and propagation length, but the spatial profiles of their field along the $z$ axis also differ. Figure 2 presents plots of the magnitudes of the Cartesian components of $\boldsymbol{e}(z)$ and $\boldsymbol{h}(z)$ as functions of $z$ for the SPP wave with $q = (1.2066 + i0.0015)k_0$, while Fig. 3 presents the same plots for the SPP wave with $q = (2.5826 + i0.0242)k_0$, when $\lambda_0 = 635$ nm. Also shown are the variations of the Cartesian components of the time-averaged Poynting vector $\boldsymbol{P}(r) = (1/2)\,\mathrm{Re}[\boldsymbol{E}(r) \times \boldsymbol{H}^*(r)]$ with $z$ on the line $\{x = 0, y = 0\}$. On the metal side, both SPP waves decay almost similarly with distance from the plane $z = 0$. However, on the other side of the interface, the SPP wave in Fig. 2 is quite weakly localized to the interface, whereas the SPP wave in Fig. 3 is very strongly localized to the interface.

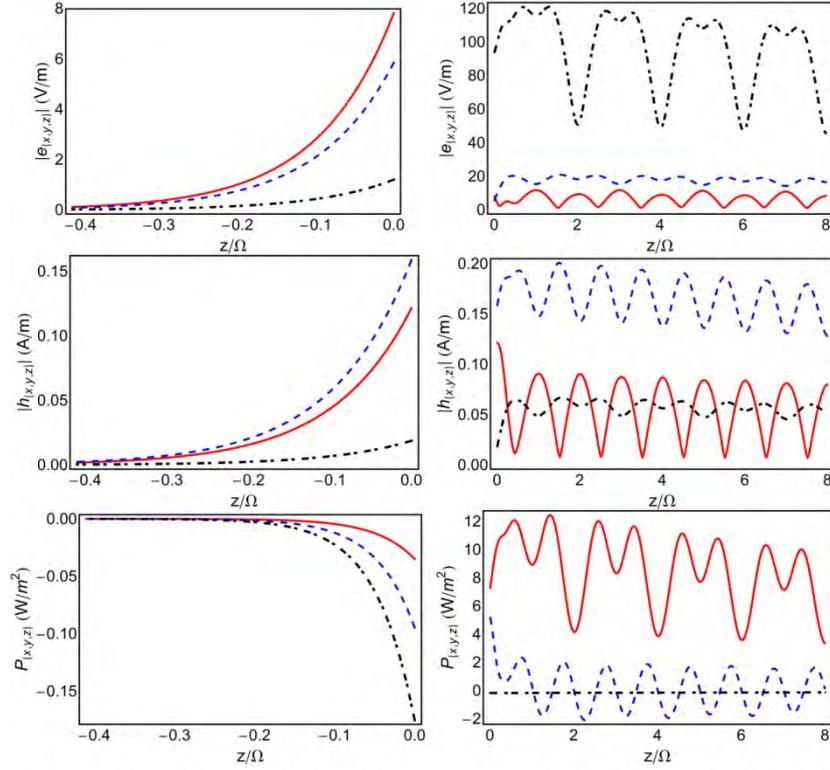

Fig. 2 Normalized variations with $z$ of the magnitudes of the components of $\boldsymbol{e}(z)$, $\boldsymbol{h}(z)$, and $\boldsymbol{P}(0,0,z)$ for an SPP wave when $\varepsilon_m = -56 + i21$, $\lambda_0 = 635$ nm, $\varepsilon_a = 2.26(1+i\delta)$, $\varepsilon_b = 3.46(-1+i\delta)$, $\varepsilon_c = 2.78(1+i\delta)$, $\delta = 0.001$, $h = +1$, $\Omega = 135$ nm, $\chi = \pi/6$, and $\psi = 0$. The $x, y,$ and $z$-components are represented by red solid, blue dashed, and black chain-dashed lines, respectively. For this SPP wave, $q = (1.2066 + i0.0015)k_0$.

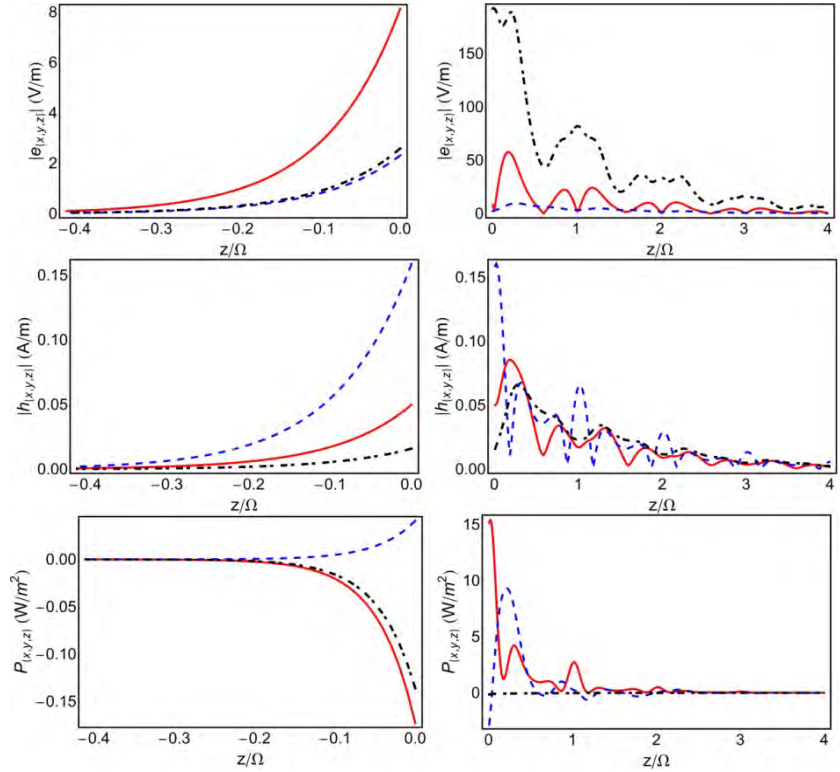

Fig. 3. Same as Fig. 2 except for the SPP wave with $q = (2.5826 + i0.0242)k_0$.

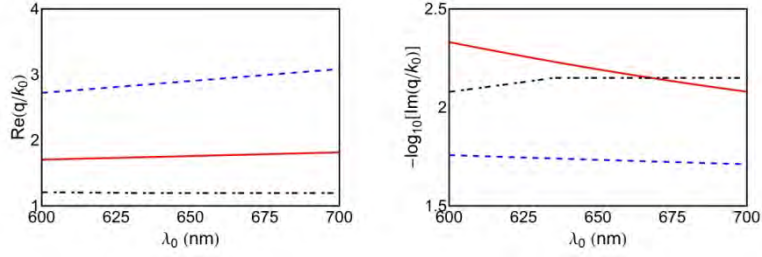

Fig. 4 Same as Fig. 1 except for $\psi = \pi/4$.

The SPP-wave-propagation phenomenon depends strongly on $\psi$. As an example, Fig. 4 presents the solutions of the dispersion equation for the same set of partnering materials as the one used for Fig. 1, except that the SPP waves propagate parallel to the unit vector $(\hat{u}_x + \hat{u}_y)/2$ in the $xy$ plane. The solutions $q$ can be organized in three branches for $\psi = \pi/4$, unlike for $\psi = 0$. Furthermore, all three branches for $\psi = \pi/4$ span the entire spectral range $\lambda_0 \in [600, 700]$ nm, but only one of the two for $\psi = 0$ does the same. Finally, the differences in the phase speeds and the propagation lengths of SPP waves on the three branches are not as wide for $\psi = \pi/4$ as for $\psi = 0$.

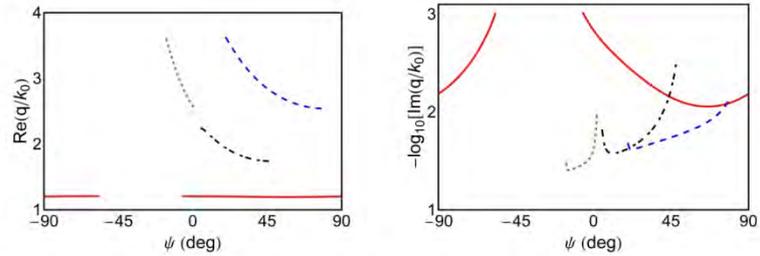

Fig. 5 Real and imaginary parts of $q$ as functions of $\psi$ when $\lambda_0 = 635$ nm, $\varepsilon_m = -56 + i21$, $\varepsilon_a = 2.26(1 + i\delta)$, $\varepsilon_b = 3.46(-1 + i\delta)$, $\varepsilon_c = 2.78(1 + i\delta)$, $\delta = 0.001$, $h = +1$, $\Omega = 135$ nm, and $\chi = 30°$.

To further illustrate the strong dependence of SPP-wave propagation on the direction of propagation in the interface plane, the real and imaginary parts of $q/k_0$ are presented in Fig. 5 as functions of $\psi \in [-90°, 90°]$ when $\lambda_0 = 635$ nm and $\chi = 30°$. The figure shows that at least one solution exists for all values of $\psi \in [-90°, -57°] \cup [-6°, 90°]$. Of course, because of chiral symmetry about the z axis the solutions of the dispersion equation are the same for $\psi$ and $\psi \pm 180°$.

The tilt angle $\chi$ affects the propagation of SPP waves significantly, as can be seen from Fig. 6 for the same parameters as for Figs. 2 and 3, except that $\chi$ is variable. No solutions were found in the range $Re(q)/k_0 \in (1, 4)$ for $\chi \in (34.9°, 42.8°) \cup (47.6°, 90°]$. Although the solutions for $\chi \in [0°, 34.9°] \cup [42.8°, 47.6°]$ can be organized in three branches, only one solution exists for much of this range and two solutions are possible only for $\chi \in [29.6°, 32.6°]$.

## 3. CONCLUDING REMARKS

Thus, the chosen interface may guide none, one, or more than one SPP waves, depending on (i) the constitutive parameters of the hyperbolic, dielectric, structurally chiral partner and (ii) the direction of propagation in the interface plane.

Although we have discussed the solutions of the canonical problem for SPP waves guided by the planar interface of a metal and a hyperbolic, dielectric, structurally chiral material, the geometric configuration of this problem is not practically implementable. For experimental excitation of SPP waves, apparatus involving prisms [3, 6, 20] and grating [21, 22] are generally used. The apparatus containing gratings are better for exciting SPP waves with high magnitudes of $Re(q)/k_0$ because prisms of very high refractive index (say, greater than 2.6) are uncommon in the spectral regime of optics. In subsequent research, we plan to examine soon the grating-coupled excitation of the SPP waves predicted in this paper.

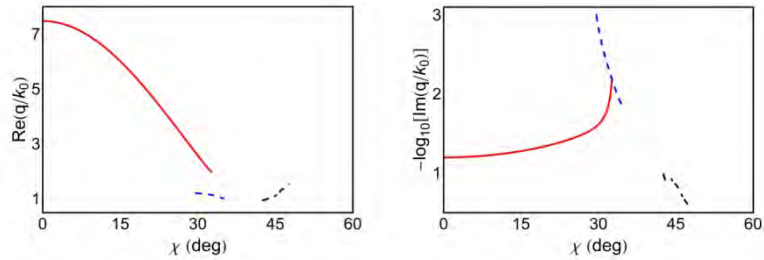

Fig. 6 Real and imaginary parts of $q$ as functions of $\chi$ when $\lambda_0 = 635$ nm, $\varepsilon_m = -56 + i21$, $\varepsilon_a = 2.26(1 + i\delta)$, $\varepsilon_b = 3.46(-1 + i\delta)$, $\varepsilon_c = 2.78(1 + i\delta)$, $\delta = 0.001$, $h = +1$, $\Omega = 135$ nm, and $\psi = 0$.